\documentclass[12pt]{article}
\newcommand{\rep}{representation}
\newcommand{\tr}{\mathop{\rm tr}}
\newcommand{\cO}{{\cal O}}
\newcommand{\half}{{1\over2}}

\newcommand{\figsize}{\small}
\renewcommand{\bar}{\overline}

\input prepictex
\input pictex
\input postpictex
\newdimen\tdim
\tdim=\unitlength
\def\stpltsmbl{\setplotsymbol ({\small .})}
\def\bsmbl{\setplotsymbol ({\Huge .})}
\def\tarrow{\arrow <5\tdim> [.3,.6]}
\def\barrow{\arrow <8\tdim> [.3,.6]}

\newcommand{\ximn}{\xi_{M/N}}

\begin{document}
\begin{titlepage}
\def\thepage {}  

\title{Dynamically broken Topcolor at Large-$N$\thanks{Research supported in
part by the
National Science Foundation
under grant number NSF-PHY/98-02709.}\,\,\thanks{HUTP-99/A042}
}

\author{
Hael Collins,\thanks{hael@pauli.harvard.edu}\,\,
Aaron Grant\thanks{grant@pauli.harvard.edu}\,\,
and Howard Georgi\thanks{georgi@physics.harvard.edu}\\
Lyman Laboratory of Physics\\
Harvard University\\
Cambridge, MA 02138}

\date{July, 1999}

\maketitle

\vspace{24pt}

\begin{abstract}

We analyze a model of dynamically broken topcolor in the limit in which the
number of colors is large. We show that the second
order nature of the phase transition, necessary for the success of topcolor
models, passes the nontrivial check of consistency with the large $N$ limit.
We also identify and discuss a class of theories that
generalizes the topcolor phenomenon to a theory with a richer structure of
fermions and global symmetries. 

\pagestyle{empty}
\end{abstract}
\end{titlepage}

\setcounter{section}{0}
\setcounter{equation}{0}

\section{Introduction\label{introduction}}

More than thirty years after Weinberg's theory of leptons~\cite{weinberg},
the nature of electroweak symmetry breaking remains obscure. Symmetry
breaking by a fundamental Higgs multiplet has been thoroughly studied in
perturbation theory. Particularly when combined with the beautiful
theoretical idea of supersymmetry, this alternative is quite attractive to
theorists~\cite{susy}. Dynamical symmetry breaking is harder to study
because it is necessarily nonperturbative. Nevertheless, this option is
highly constrained by precise data on the properties of the $W$ and $Z$
bosons, and by more general considerations such as the suppression of
flavor changing neutral currents and the apparent existence of small
neutrino masses~\cite{technicolor}. Topcolor~\cite{topcolor} is one of the
few dynamical schemes that still seems viable. In topcolor, the Higgs
multiplet exists, but is a composite state of fermions (including the left
handed $t$ quark) that carry the strong topcolor interaction. It is
necessary in this scheme that the topcolor gauge symmetry is spontaneously
broken down to ordinary color (leaving massive, strongly interacting
``colorons'' that contribute to the binding of the Higgs). A critical
component of topcolor is the assumption that the mass squared of the
composite Higgs can be tuned to be very small compared to the scale of the
topcolor dynamics. This is certainly the case if the electroweak symmetry
breaking in the model is a second order phase transition.

Our goal in this note is not to build realistic models, but rather to
develop tools that may be useful for model building. Most previous studies
of topcolor models have concentrated on the gauge structure and the
couplings of the colorons to the $t$ and other quarks. Here, we focus
instead on topcolor breaking. We will study a class of theories in which
this symmetry breaking is itself dynamical. We analyze these in the limit
in which the number of colors is large. We show that the second order
nature of the phase transition, necessary for the success of topcolor
models, passes the nontrivial check of consistency with the large $N$
limit~\cite{eichtenpreskill}. We also identify and discuss a class of
theories that generalizes the topcolor phenomenon to a theory with a richer
structure of fermions and global symmetries.

\section{The simple Moose\label{moose}}

We begin by considering not a topcolor model, but something simpler --- an
$SU(N)\times SU(M)$ gauge theory with three multiplets of massless
fermions and an $SU(M)\times SU(N)$ global
symmetry.\footnote{Here and below we indicate gauge symmetries with a
subscript $g$.} The theory is most usefully described by the
``Moose''~\cite{mooses} diagram in figure~\ref{fig-1}.
{\figsize\begin{figure}[htb]
$$\beginpicture
\setcoordinatesystem units <\tdim,\tdim>
\stpltsmbl
\circulararc 360 degrees from 215 0 center at 200 0
\circulararc 360 degrees from 115 0 center at 100 0
\put {$M$} at 0 0
\put {$M$} at 200 0
\put {$N$} at 100 0
\put {$N$} at 300 0
\put {$A$} at 50 -15 
\put {$B$} at 150 -15 
\put {$C$} at 250 -15 
\tarrow from 15 0 to 50 0
\plot 50 0 85 0 /
\tarrow from 115 0 to 150 0
\plot 150 0 185 0 /
\tarrow from 215 0 to 250 0
\plot 250 0 285 0 /
\setdashes
\put {$SU(M)$} at 0 -50
\tarrow from 0 -40 to 0 -10 
\put {$SU(M)_g$} at 200 -50
\tarrow from 200 -40 to 200 -20 
\put {$SU(N)$} at 300 -50
\tarrow from 300 -40 to 300 -10 
\put {$SU(N)_g$} at 100 -50
\tarrow from 100 -40 to 100 -20 
\endpicture$$
\caption{\figsize\label{fig-1}} \sf The basic Moose --- in this model, we
expect
no phase transition as a function of the ratio of coupling strengths.
\end{figure}}
The left-handed (LH) fermions are represented by the solid directed lines,
labeled $A$, $B$ and $C$. The gauge groups, all $SU(K)$ for some $K$, are
represented by circles with the value of $K$ inside. Lines that do not end
in circles transform under a global flavor $SU(K)$ symmetry, again with the
value of $K$ indicated on the diagram. The Moose encodes the
transformation properties of the LH fermions under the various gauge and
global symmetries of the model, in this case,
\begin{equation}
SU(M)
\times SU(N)_g
\times SU(M)_g
\times SU(N).
\end{equation}
The fermion associated with a directed line in Moose notation transforms
like the defining \rep\ if the arrow leads away from the label for the
group, and like the complex conjugate of the defining \rep\ if the arrow
leads toward to the group label. Thus
\begin{itemize}
\item $A$ transforms like $(M,\bar N,1,1)$
\item $B$ transforms like $(1,N,\bar M,1)$
\item $C$ transforms like $(1,1,M,\bar N)$
\end{itemize}
This theory depends on two parameters --- the gauge couplings (or more
properly, because of dimensional transmutation~\cite{cw}, the $\Lambda$
parameters) of the two gauge groups. We have argued
elsewhere~\cite{mooses} that independent of the ratio of the gauge
couplings the model has an unbroken $SU(M)\times SU(N)$ global symmetry and
a multiplet of massless LH fermions transforming like $(M,\bar N)$
saturating the global anomaly conditions~\cite{thooft}.\footnote{We will
not review these arguments here, because we will be going through similar
arguments in considerable detail in section~\ref{knotzero}.} If either of
the gauge couplings is much smaller than the other, we are very confident
that this is the case because the strong interactions are QCD-like. And
there is no reason to expect any phase transition as a function of the
ratio of couplings~\cite{complementarity}. If the couplings are of the same
order of magnitude, it makes sense to refer to the massless fermions as
composites. It is this region that we will find particularly interesting.
{\figsize\begin{figure}[htb]
$$\beginpicture
\setcoordinatesystem units <\tdim,\tdim>
\stpltsmbl
\circulararc 360 degrees from 215 0 center at 200 0
\circulararc 360 degrees from 115 0 center at 100 0
\put {$M$+$k$~~} at 0 0
\put {$M$} at 200 0
\put {$N$} at 100 0
\put {$k$} at 100 100
\put {$N$} at 300 0
\tarrow from 100 15 to 100 50
\plot 100 50 100 85 /
\tarrow from 15 0 to 50 0
\plot 50 0 85 0 /
\tarrow from 115 0 to 150 0
\plot 150 0 185 0 /
\tarrow from 215 0 to 250 0
\plot 250 0 285 0 /
\endpicture$$
\caption{\figsize\label{fig-2}} \sf The Moose with a small side chain --- in
this
model, we expect a second order phase transition as a function of the ratio of
coupling strengths. \end{figure}}
{\figsize\begin{figure}[htb]
$$\beginpicture
\setcoordinatesystem units <\tdim,\tdim>
\stpltsmbl
\circulararc 360 degrees from 215 0 center at 200 0
\circulararc 360 degrees from 115 0 center at 100 0
\put {$M$+$k$~~} at 0 0
\put {$M$} at 200 0
\put {$\ell$} at 200 100
\put {$N$} at 100 0
\put {$k$} at 100 100
\put {~~$N$+$\ell$} at 300 0
\tarrow from 200 85 to 200 50
\plot 200 50 200 15 / 
\tarrow from 100 15 to 100 50
\plot 100 50 100 85 /
\tarrow from 15 0 to 50 0
\plot 50 0 85 0 /
\tarrow from 115 0 to 150 0
\plot 150 0 185 0 /
\tarrow from 215 0 to 250 0
\plot 250 0 285 0 /
\endpicture$$
\caption{\figsize\label{fig-3}} \sf The Moose with two small side chains ---
in this
model, we expect (generically) two second order phase transitions as a
function of
the ratio of coupling strengths. \end{figure}}

In this note, we will study the limit of this theory in which $N$ and $M$
go to infinity in fixed ratio, with gauge couplings scaling like
$1/\sqrt{N}$. We had hoped to be able contribute some additional evidence
for the conjecture that there is no phase transition in the theory. We have
been frustrated in this attempt, and will discuss some of the reasons in
appendix~\ref{condensates}. In what follows, we will simply assume that
this conjecture is true, and remains true in the limit. We will consider
instead the phase structure of related and more interesting models in the
same large $N$ limit. In particular, we will discuss below the models
described by the Mooses in figures~\ref{fig-2} and \ref{fig-3}, and the
suggest that phase transitions that must occur in these models are second
order in the limit that $N$ and $M$ go to infinity with $k$ and $\ell$
fixed. We will show that the states that exist in the large $N$ limit are
just those required for the consistency of second order phase transitions
in these models. The Moose in figure~\ref{fig-2} is the business end of
model of a dynamically broken topcolor~\cite{topcolor,seesaw,seesaw2}, with
only the topcolor (and not the color) gauge symmetry, and without the
fields that do not carry topcolor or cause its breakdown. The Moose in
figure~\ref{fig-3} is a nontrivial generalization of the topcolor
phenomenon that is only possible in a theory in which topcolor is
dynamically broken.

\section{Large $N$ and $M$ for $k=\ell=0$\label{largenandm}}

Here we discuss the limit of the Moose in figure~\ref{fig-1} in which $N$
and $M$ go to infinity with the $\Lambda$ parameters of the gauge groups
fixed. The first thing to note is that we must take $N$ and $M$ to infinity
at the same rate. Otherwise, one or the other of the gauge couplings would
lose asymptotic freedom. Thus we assume that the ratio $M/N$ is fixed in
the limit. The limit thus depends both on $M/N$ and on
$\Lambda_M/\Lambda_N$, but we will usually regard $M/N$ as fixed and
concentrate on the dependence on $\Lambda_M/\Lambda_N$. We will refer to
all of this simply as the large $N$ limit.

For finite $M$ and $N$, the model has a single anomaly free $U(1)$
symmetry. Two combinations of the three classical $U(1)$s are broken by
the anomalies of the two gauge groups. However, in the large $N$ limit, the
gauge anomalies disappear and the classical $U(1)$ symmetries are
restored. These $U(1)$ symmetries will not play a crucial role in anything
we do, and we will usually ignore them completely.

Let us first consider vacuum to vacuum amplitudes in this large $N$ limit.
This is a matrix model, in which we can think of each of the fermions and
the gauge bosons as represented by a pair of oppositely directed lines. For
the gauge bosons, the two lines refer to the same gauge index, while for
the fermions, the two lines are associated with different gauge or global
indices. But since $M$ and $N$ are going to infinity at the same rate,
these differences are irrelevant to the diagrammatics. The leading vacuum
to vacuum graphs are therefore those that can be drawn on the surface of a
sphere, like pure gluon graphs in QCD.

There are three kinds of interesting bound states in this model that carry the
flavor symmetries:
\begin{description}
\item[mesons which transform like adjoints under
the $SU(M)$ flavor symmetry] --- these are built of $A\bar A$ plus
flavor singlet combinations of the fermions;
\item[mesons which transform like adjoints under
the $SU(N)$ flavor symmetry] --- these are built of $C\bar C$ plus
flavor singlet combinations of the fermions;
\item[baryons which transform like $(M,\bar N)$ under
the $SU(M)\times SU(N)$ flavor symmetry] --- these are built of $ABC$
plus flavor singlet combinations of the fermions.
\end{description}
Because this is a matrix model, that is, because the number of flavors is
going to infinity like the number of colors, we cannot classify the states
in terms of their fermion constituents. Arbitrary numbers of all varieties
of fermion-antifermion pairs appear in the wave-function in the same order
in $N$. In addition, the bound states are not free in the large $N$ limit
as they are in large $N$ QCD with a fixed number of flavors. The couplings
of the bound states to one another go to zero, but because the number of
bound states is going to infinity, the interactions cannot be
neglected.\footnote{In other words, the effective theory describing the
bound states is a matrix model in which the number of flavors is going to
infinity.} However, the flavor structure of the graphs contributing to the
propagation and interactions of the particles that carry flavor is still
very simple. The leading graphs are those in which all the flavor
dependence is associated with the lines running around the outside of a
planar diagram. We will now illustrate all this in more detail using the
following notation for the fermion lines propagating to the right:
\begin{equation}
\begin{tabular}{cl}
$A$ &{\beginpicture
\setcoordinatesystem units <\tdim,\tdim>
\stpltsmbl
\barrow from 48 0 to 53 0
\bsmbl
\setdots
\plot 0 0 100 0 /
\linethickness=0pt \putrule from 0 0 to 100 0
\endpicture}\\
$B$ &{\beginpicture
\setcoordinatesystem units <\tdim,\tdim>
\stpltsmbl
\barrow from 52 0 to 47 0
\plot 0 0 100 0 /
\linethickness=0pt \putrule from 0 0 to 100 0
\endpicture}\\
$C$ &{\beginpicture
\setcoordinatesystem units <\tdim,\tdim>
\stpltsmbl
\barrow from 48 0 to 53 0
\setdashes
\plot 0 0 100 0 /
\linethickness=0pt \putrule from 0 0 to 100 0
\endpicture}
\end{tabular}
\label{lines}
\end{equation}
Note that we have defined the direction of the arrow on the $B$ to be
opposite to that of $A$ and $C$ so that the lines with their arrows in the
same direction are carrying the same kind of gauge charge ($A$ is an $\bar
N$ while $B$ is an $N$ of $SU(N)$ and $C$ is an $M$ while $B$ is an $\bar
M$ of $SU(M)$). We begin by showing a series of figures that illustrate
meson and baryon propagation. We will discuss interactions in the next
section. 
{\figsize\begin{figure}[htb]
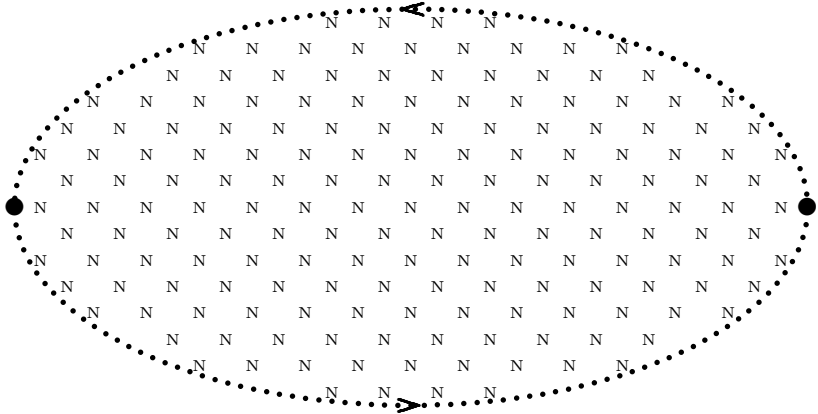

$$\beginpicture
\setcoordinatesystem units <\tdim,\tdim>
\stpltsmbl
\barrow from -2 75 to -3 75
\barrow from 2 -75 to 3 -75
\bsmbl
\setdots
\ellipticalarc axes ratio 2:1 360 degrees from 150 0 center at 0 0
\setshadesymbol <0pt,,,> ({\tiny N} <0pt,0pt>)
\setshadegrid span <10\tdim>
\vshade -150 0 0
-140 -27.0 27.0 -130 -37.4 37.4 -120 -45.0 45.0 -110 -51.0 51.0 -100 -56.0
56.0 -90 -60.0 60.0 -80 -63.5 63.5 -70 -66.5 66.5 -60 -68.5 68.5 -50 -70.5
70.5 -40 -72.5 72.5 -30 -73.5 73.5 -20 -74.5 74.5 -10 -75.0 75.0 0
-75.0 75.0 10 -75.0 75.0 20 -74.5 74.5 30 -73.5 73.5 40 -72.5 72.5 50 -70.5
70.5 60 -68.5 68.5 70 -66.5 66.5 80 -63.5 63.5 90 -60.0 60.0 100 -56.0
56.0 110 -51.0 51.0 120 -45.0 45.0 130 -37.4 37.4 140 -27.0 27.0 150 0 0 /
\put {\Large$\bullet$} at -150 0
\put {\Large$\bullet$} at 150 0
\linethickness=0pt
\putrule from -150 0 to 150 0
\putrule from 0 -77 to 0 77
\endpicture$$
\caption{\figsize\sf\label{fig-p1} A class of diagrams contributing to the
propagation of $A\bar A$ adjoint mesons in leading order in large $N$. The
{\Large$\bullet$}s indicate an $\bar A T_a A$ interpolating field. The
small Ns indicate arbitrary planar dressing of $SU(N)$ gluons.
}\end{figure}}

Figure~\ref{fig-p1} illustrates the typical planar graphs that contribute to
meson
propagation in a large $N$ gauge theory with a finite number of flavors, in
this case, illustrated for the $A\bar A$ mesons that are bound by the $SU(N)$
gauge interactions and carry $SU(M)$ flavor. The
small Ns indicate arbitrary planar dressing of $SU(N)$ gluons. There are
corresponding graphs
with a $C$ loop running around the outside and dressed with $SU(M)$ gluons
that contribute to the propagation of $C\bar C$ mesons
{\figsize\begin{figure}[htb]
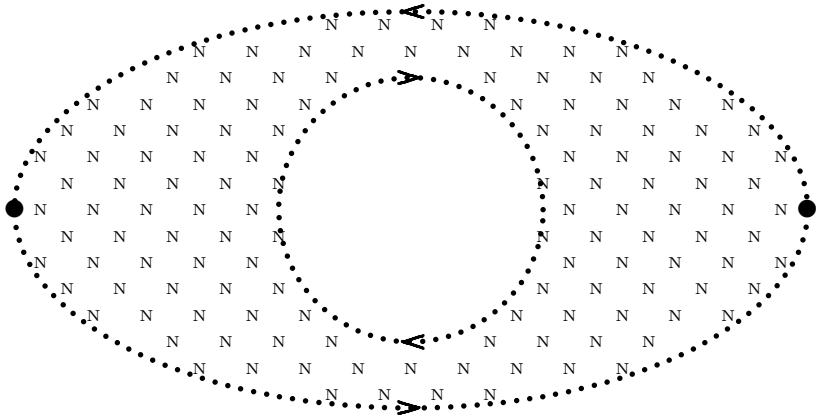

$$\beginpicture
\setcoordinatesystem units <\tdim,\tdim>
\stpltsmbl
\barrow from -2 75 to -3 75
\barrow from 2 -75 to 3 -75
\barrow from -2 -50 to -3 -50
\barrow from 2 50 to 3 50
\bsmbl
\setdots
\ellipticalarc axes ratio 2:1 360 degrees from 150 0 center at 0 0
\circulararc 360 degrees from 50 0 center at 0 0
\setshadesymbol <0pt,,,> ({\tiny N} <0pt,0pt>)
\setshadegrid span <10\tdim>
\vshade -150 0 0
-140 -27.0 27.0 -130 -37.4 37.4 -120 -45.0 45.0 -110 -51.0 51.0 -100 -56.0
56.0 -90 -60.0 60.0 -80 -63.5 63.5 -70 -66.5 66.5 -60 -68.5 68.5 -50 -70.5
70.5 /
\vshade 50 -70.5
70.5 60 -68.5 68.5 70 -66.5 66.5 80 -63.5 63.5 90 -60.0 60.0 100 -56.0
56.0 110 -51.0 51.0 120 -45.0 45.0 130 -37.4 37.4 140 -27.0 27.0 150 0 0 /
\vshade -50 0 70.5 -40 30.0 72.5 -30 40.0 73.5 -20 45.8 74.5 -10 49.0 75.0 0
50.0 75.0 10 49.0 75.0 20 45.8 74.5 30 40.0 73.5 40 30.0 72.5 50 0 70.5 /
\vshade -50 -70.710 0 -40 -72.285 -30.000 -30 -73.485 -40.000 -20 -74.330
-45.826 -10 -74.835 -48.990 0 -75.000 -50.000 10 -74.835 -48.990 20 -74.330
-45.826 30 -73.485 -40.000 40 -72.285 -30.000 50 -70.710 0 /
\put {\Large$\bullet$} at -150 0
\put {\Large$\bullet$} at 150 0
\linethickness=0pt
\putrule from -150 0 to 150 0
\putrule from 0 -77 to 0 77
\endpicture$$
\caption{\figsize\sf\label{fig-p2} Another class of diagrams contributing to
the
propagation of $A\bar A$ adjoint mesons in leading order in large $N$. The
internal $A$ loop produces a hole in the planar diagram, but it involves a sum
over all $SU(M)$ flavors and thus does not affect the flavor structure. These
graphs are of order $M/N=\cO(1)$ compared to those of
figure~\protect\ref{fig-p1}.
}\end{figure}}

Because the number of flavors grows with $N$, internal loops are not
suppressed, as illustrated in figure~\ref{fig-p2}. These graphs are
suppressed by the large $N$ gluon couplings, but enhanced by the $M$-flavor
sum in the internal loop, so they are of order $M/N=\cO(1)$ compared to
those of figure~\protect\ref{fig-p1}. Arbitrary numbers of loops may
appear, so that planar graphs for a large number of flavors may actually
resemble a slice of Swiss cheese. Note that there is an orientation to the
planar gluon graphs determined by the direction of the outside loop. If the
external arrows are on the right of the planar gluon dressing as in this
case, then to contribute in leading order, all internal $A$ lines must have
their arrows on the right in the same way.\footnote{You can see this in
detail by putting in some gluon lines and redrawing the diagram in the
double-line notation.} Thus internal (empty) $A$ loops must circulate in
the opposite direction from the external loop. Graphs in which an internal
$A$ loop goes around in the other direction are non-leading in $N$. The
internal $A$ loop involves a sum over all the $SU(M)$ flavors (that is why
it contributes in leading order), and thus it has no affect of the flavor
structure of the graph, which is entirely associated with the outside
line. We will return to this when we discuss interactions below.

{\figsize\begin{figure}[htb]
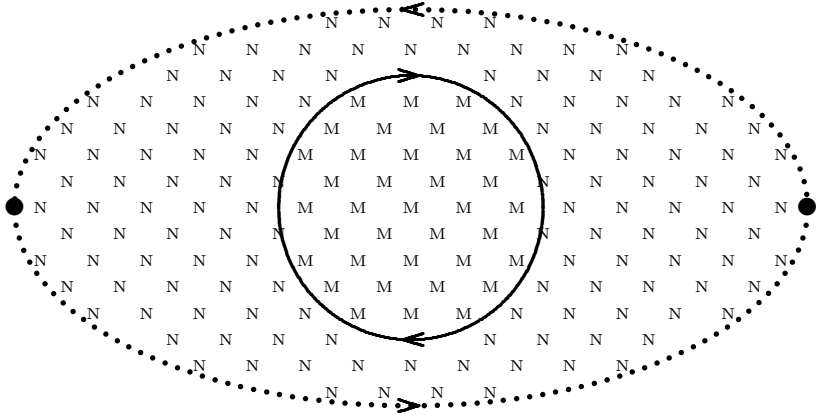

$$\beginpicture
\setcoordinatesystem units <\tdim,\tdim>
\stpltsmbl
\barrow from -2 75 to -3 75
\barrow from 2 -75 to 3 -75
\barrow from -2 -50 to -3 -50
\barrow from 2 50 to 3 50
\circulararc 360 degrees from 50 0 center at 0 0
\bsmbl
\setdots
\ellipticalarc axes ratio 2:1 360 degrees from 150 0 center at 0 0
\setshadesymbol <0pt,,,> ({\tiny N} <0pt,0pt>)
\setshadegrid span <10\tdim>
\vshade -150 0 0
-140 -27.0 27.0 -130 -37.4 37.4 -120 -45.0 45.0 -110 -51.0 51.0 -100 -56.0
56.0 -90 -60.0 60.0 -80 -63.5 63.5 -70 -66.5 66.5 -60 -68.5 68.5 -50 -70.5
70.5 /
\vshade 50 -70.5
70.5 60 -68.5 68.5 70 -66.5 66.5 80 -63.5 63.5 90 -60.0 60.0 100 -56.0
56.0 110 -51.0 51.0 120 -45.0 45.0 130 -37.4 37.4 140 -27.0 27.0 150 0 0 /
\vshade -50 0 70.5 -40 30.0 72.5 -30 40.0 73.5 -20 45.8 74.5 -10 49.0 75.0 0
50.0 75.0 10 49.0 75.0 20 45.8 74.5 30 40.0 73.5 40 30.0 72.5 50 0 70.5 /
\vshade -50 -70.710 0 -40 -72.285 -30.000 -30 -73.485 -40.000 -20 -74.330
-45.826 -10 -74.835 -48.990 0 -75.000 -50.000 10 -74.835 -48.990 20 -74.330
-45.826 30 -73.485 -40.000 40 -72.285 -30.000 50 -70.710 0 /
\setshadesymbol <0pt,,,> ({\tiny M} <0pt,0pt>)
\vshade -50 0 0 -40 -30.000 30.000 -30 -40.000 40.000 -20 -45.826 45.826 -10
-48.990 48.990 0 -50.000 50.000 10 -48.990 48.990 20 -45.826 45.826 30 -40.000
40.000 40 -30.000 30.000 50 0 0 /
\put {\Large$\bullet$} at -150 0
\put {\Large$\bullet$} at 150 0
\linethickness=0pt
\putrule from -150 0 to 150 0
\putrule from 0 -77 to 0 77
\endpicture$$
\caption{\figsize\sf\label{fig-p3} Another class of diagrams contributing to
the
propagation of $A\bar A$ adjoint mesons in leading order in large $N$. The $B$
loop encloses a region full of small Ms indicating arbitrary planar dressing
of $SU(M)$ gluons. Again the flavor structure is unaffected.
}\end{figure}}
The prominence of graphs like those in figure~\ref{fig-p2} is related to the
fact that the meson resonances in a model like this are not narrow, but have
unsuppressed decays into multi-meson states~\cite{veneziano}. We will come
back to this also in
section~\ref{interactions}.

{\figsize\begin{figure}[htb]
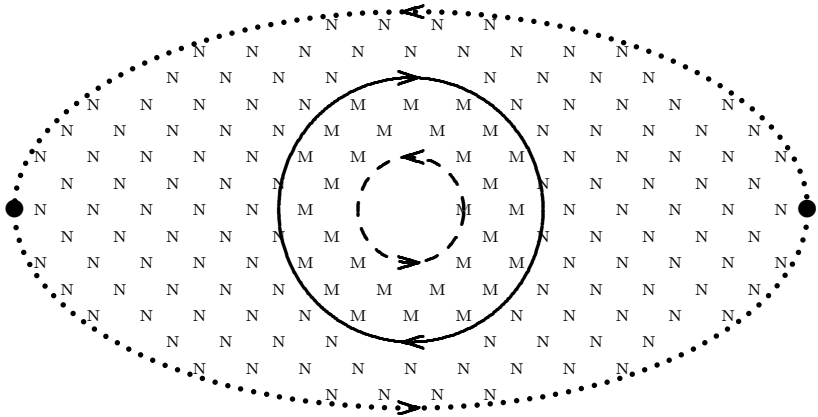

$$\beginpicture
\setcoordinatesystem units <\tdim,\tdim>
\stpltsmbl
\barrow from -2 75 to -3 75
\barrow from 2 -75 to 3 -75
\barrow from -2 -50 to -3 -50
\barrow from 2 50 to 3 50
\barrow from -2 20 to -3 20
\barrow from 2 -20 to 3 -20
\circulararc 360 degrees from 50 0 center at 0 0
\setdashes
\circulararc 360 degrees from 20 0 center at 0 0
\bsmbl
\setdots
\ellipticalarc axes ratio 2:1 360 degrees from 150 0 center at 0 0
\setshadesymbol <0pt,,,> ({\tiny N} <0pt,0pt>)
\setshadegrid span <10\tdim>
\vshade -150 0 0
-140 -27.0 27.0 -130 -37.4 37.4 -120 -45.0 45.0 -110 -51.0 51.0 -100 -56.0
56.0 -90 -60.0 60.0 -80 -63.5 63.5 -70 -66.5 66.5 -60 -68.5 68.5 -50 -70.5
70.5 /
\vshade 50 -70.5
70.5 60 -68.5 68.5 70 -66.5 66.5 80 -63.5 63.5 90 -60.0 60.0 100 -56.0
56.0 110 -51.0 51.0 120 -45.0 45.0 130 -37.4 37.4 140 -27.0 27.0 150 0 0 /
\vshade -50 0 70.5 -40 30.0 72.5 -30 40.0 73.5 -20 45.8 74.5 -10 49.0 75.0 0
50.0 75.0 10 49.0 75.0 20 45.8 74.5 30 40.0 73.5 40 30.0 72.5 50 0 70.5 /
\vshade -50 -70.710 0 -40 -72.285 -30.000 -30 -73.485 -40.000 -20 -74.330
-45.826 -10 -74.835 -48.990 0 -75.000 -50.000 10 -74.835 -48.990 20 -74.330
-45.826 30 -73.485 -40.000 40 -72.285 -30.000 50 -70.710 0 /
\setshadesymbol <0pt,,,> ({\tiny M} <0pt,0pt>)
\vshade -50 0 0 -40 -30.000 30.000 -30 -40.000 40.000 -20 -45.826 45.826 /
\vshade 20 -45.826 45.826 30 -40.000 40.000 40 -30.000 30.000 50 0 0 /
\vshade -20 0 45.8 -10 17.3 49.0 0 20.0 50.0 10 17.3 49.0 20 0 45.8 /
\vshade -20 -45.8 0 -10 -49.0 -17.3 0 -50.0 -20.0 10 -49.0 -17.3 20 -45.8 0 /
\put {\Large$\bullet$} at -150 0
\put {\Large$\bullet$} at 150 0
\linethickness=0pt
\putrule from -150 0 to 150 0
\putrule from 0 -77 to 0 77
\endpicture$$
\caption{\figsize\sf\label{fig-p4} Another class of diagrams contributing to
the
propagation of $A\bar A$ adjoint mesons in leading order in large $N$. The $C$
loop produces hole in a region dressed
with $SU(M)$ gluons. As always, the flavor structure comes entirely from the
loop on the outside.
}\end{figure}}
We may also have an internal $B$ loop, as illustrated in figure~\ref{fig-p3}.
Because the $B$ carries both the gauge $SU(N)$ and $SU(M)$, it interacts with
$N$-gluons on one side and with $M$-gluons on the other, and thus forms a
boundary between regions in the planar diagram, as shown. Note that the
$M$-region has the opposite orientation, in the sense that the arrows are on
the left of the gluon dressing.

{\figsize\begin{figure}[htb]
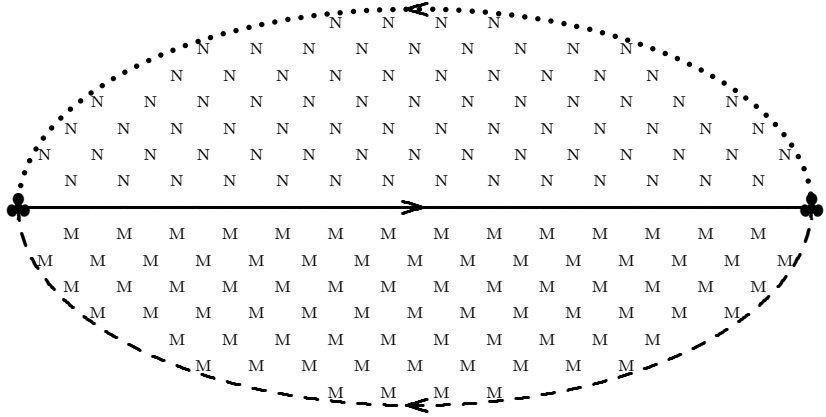

$$\beginpicture
\setcoordinatesystem units <\tdim,\tdim>
\stpltsmbl
\barrow from -2 75 to -3 75
\barrow from -2 -75 to -3 -75
\barrow from 2 0 to 3 0
\plot -150 0 150 0 /
\setdashes
\ellipticalarc axes ratio 2:1 180 degrees from -150 0 center at 0 0
\bsmbl
\setdots
\ellipticalarc axes ratio 2:1 180 degrees from 150 0 center at 0 0
\setshadesymbol <0pt,,,> ({\tiny N} <0pt,0pt>)
\setshadegrid span <10\tdim>
\vshade -150 0 0 -140 0 27.0 -130 0 37.4 -120 0 45.0 -110 0 51.0 -100 0 56.0
-90 0 60.0 -80 0 63.5 -70 0 66.5 -60 0 68.5 -50 0 70.5 -40 0 72.5 -30 0 73.5
-20 0 74.5 -10 0 75.0 0 0 75.0 10 0 75.0 20 0 74.5 30 0 73.5 40 0 72.5 50 0
70.5 60 0 68.5 70 0 66.5 80 0 63.5 90 0 60.0 100 0 56.0 110 0 51.0 120 0 45.0
130 0 37.4 140 0 27.0 150 0 0 /
\setshadesymbol <0pt,,,> ({\tiny M} <0pt,0pt>)
\vshade -150 0 0 -140 -27.0 0 -130 -37.4 0 -120 -45.0 0 -110 -51.0 0 -100
-56.0 0 -90 -60.0 0 -80 -63.5 0 -70 -66.5 0 -60 -68.5 0 -50 -70.5 0 -40 -72.5
0 -30 -73.5 0 -20 -74.5 0 -10 -75.0 0 0 -75.0 0 10 -75.0 0 20 -74.5 0 30 -73.5
0 40 -72.5 0 50 -70.5 0 60 -68.5 0 70 -66.5 0 80 -63.5 0 90 -60.0 0 100 -56.0
0 110 -51.0 0 120 -45.0 0 130 -37.4 0 140 -27.0 0 150 0 0 /
\put {$\clubsuit$} at -150 0
\put {$\clubsuit$} at 150 0
\linethickness=0pt
\putrule from -150 0 to 150 0
\putrule from 0 -77 to 0 77
\endpicture$$
\caption{\figsize\sf\label{fig-p5} A class of diagrams contributing to the
propagation of $ABC$ baryons in leading order in large $N$. The
{$\clubsuit$}s are interpolating fields for the baryon and antibaryon.
}\end{figure}}
A $C$ loop can produce a hole in a regions dressed by $M$-gluons, as
illustrated in figure~\ref{fig-p4}. In the internal $C$ loop, as in the $B$
loop, the arrows are on the left of the $M$ region. These diagrams are related
to decays of meson resonances into baryon-antibaryon states.

{\figsize\begin{figure}[htb]
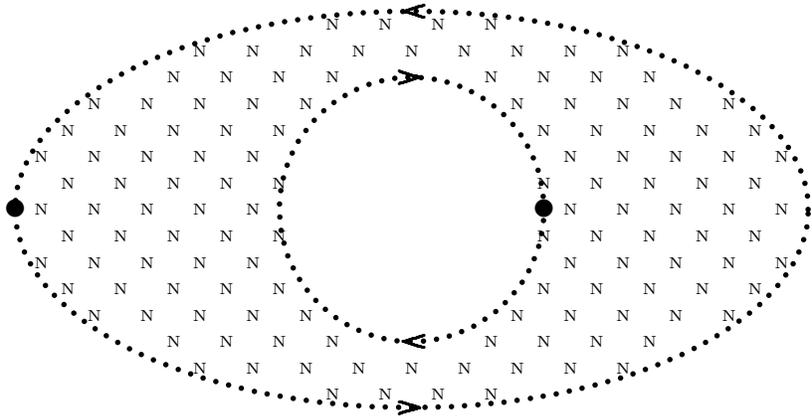

$$\beginpicture
\setcoordinatesystem units <\tdim,\tdim>
\stpltsmbl
\barrow from -2 75 to -3 75
\barrow from 2 -75 to 3 -75
\barrow from -2 -50 to -3 -50
\barrow from 2 50 to 3 50
\bsmbl
\setdots
\ellipticalarc axes ratio 2:1 360 degrees from 150 0 center at 0 0
\circulararc 360 degrees from 50 0 center at 0 0
\setshadesymbol <0pt,,,> ({\tiny N} <0pt,0pt>)
\setshadegrid span <10\tdim>
\vshade -150 0 0
-140 -27.0 27.0 -130 -37.4 37.4 -120 -45.0 45.0 -110 -51.0 51.0 -100 -56.0
56.0 -90 -60.0 60.0 -80 -63.5 63.5 -70 -66.5 66.5 -60 -68.5 68.5 -50 -70.5
70.5 /
\vshade 50 -70.5
70.5 60 -68.5 68.5 70 -66.5 66.5 80 -63.5 63.5 90 -60.0 60.0 100 -56.0
56.0 110 -51.0 51.0 120 -45.0 45.0 130 -37.4 37.4 140 -27.0 27.0 150 0 0 /
\vshade -50 0 70.5 -40 30.0 72.5 -30 40.0 73.5 -20 45.8 74.5 -10 49.0 75.0 0
50.0 75.0 10 49.0 75.0 20 45.8 74.5 30 40.0 73.5 40 30.0 72.5 50 0 70.5 /
\vshade -50 -70.710 0 -40 -72.285 -30.000 -30 -73.485 -40.000 -20 -74.330
-45.826 -10 -74.835 -48.990 0 -75.000 -50.000 10 -74.835 -48.990 20 -74.330
-45.826 30 -73.485 -40.000 40 -72.285 -30.000 50 -70.710 0 /
\put {\Large$\bullet$} at -150 0
\put {\Large$\bullet$} at 50 0
\linethickness=0pt
\putrule from -150 0 to 150 0
\putrule from 0 -77 to 0 77
\endpicture$$
\caption{\figsize\sf\label{fig-p6} A class of diagrams contributing to the
propagation of $A\bar A$ singlet mesons in leading order in large $N$.
}\end{figure}}
Finally, the propagation of baryons is described by diagrams like that
illustrated in figure~\ref{fig-p5}. Of course, there are more complicated
diagrams in which both the $N$ and $M$ regions are full of holes!

The graphs contributing to the propagation of flavor singlet
particles are more complicated, because the singlets can couple to the fermion
loops inside the planar diagram. These graphs vanish for flavor adjoint
mesons, but are leading in $N$ for the flavor singlet case. A simple example
is shown in figure~\ref{fig-p6}.

\section{Interactions in Large $N$\label{interactions}}

{\figsize\begin{figure}[htb]
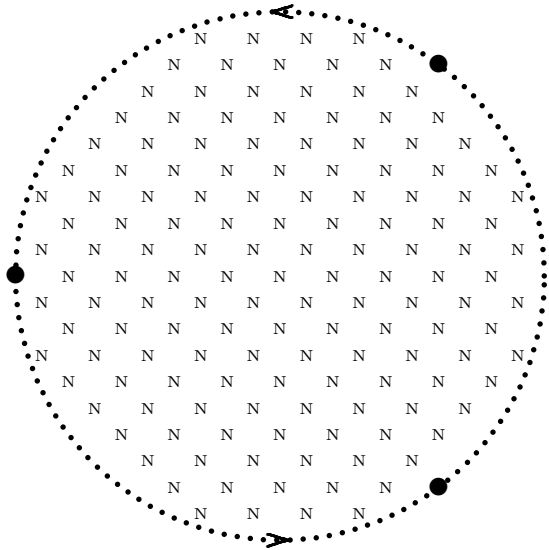

$$\beginpicture
\setcoordinatesystem units <\tdim,\tdim>
\stpltsmbl
\barrow from -2 100 to -3 100
\barrow from 2 -100 to 3 -100
\bsmbl
\setdots
\circulararc 360 degrees from 100 0 center at 0 0
\setshadesymbol <0pt,,,> ({\tiny N} <0pt,0pt>)
\setshadegrid span <10\tdim>
\vshade -100 0 0 -90 -43.6 43.6 -80 -60.0 60.0 -70 -71.4 71.4 -60 -80.0 80.0
-50 -86.6 86.6 -40 -91.7 91.7 -30 -95.4 95.4 -20 -98.0 98.0 -10 -99.5 99.5 0
-100. 100. 10 -99.5 99.5 20 -98.0 98.0 30 -95.4 95.4 40 -91.7 91.7 50 -86.6
86.6 60 -80.0 80.0 70 -71.4 71.4 80 -60.0 60.0 90 -43.6 43.6 100 0 0 /
\put {\Large$\bullet$} at -100 0
\put {\Large$\bullet$} at 60 80
\put {\Large$\bullet$} at 60 -80
\linethickness=0pt
\putrule from -150 0 to 150 0
\putrule from 0 -100 to 0 100
\endpicture$$
\caption{\figsize\sf\label{fig-i1} A class of diagrams contributing to
the interactions
of three $A\bar A$ adjoint mesons in leading order in large $N$. These graphs
vanish like $1/\sqrt N$ as $N\rightarrow\infty$, but the large number of meson
states sometimes compensates for this suppression so the interactions cannot be
ignored. Multi-meson interactions arise when more {\Large$\bullet$}s appear on
the external $A$ loop.}\end{figure}}
In this section, we will show that the leading interactions between the bound
states in the theory of figure~\ref{fig-1} are associated with graphs which,
like the propagator graphs in the previous section, are planar and have all
their flavor structure on an external loop.
It will simplify the discussion of interactions if we begin by establishing an
important general principle.
{\bf Interactions in leading order involve only a finite number of flavors at
a time.} Interactions of states
involving many flavors (like singlet states) are suppressed because of large
$N$ counting. This may seem counter-intuitive, because the states in which all
the flavors participate
can give rise to large counting factors from loops. But these
factors also appear in the propagator, and so are explicitly divided out in
the wave-function renormalization factors. When the interpolating fields for
these states then appear in more complicated graphs, the wave function
renormalization gives an additional suppression. One way of saying this is
that the interactions of singlets and other states involving order $N$ flavors
are suppressed by counting factors not only because the number of colors is
large, but also because the number of flavors is large. 

Let us illustrate this in a simple example. Compare the 3-meson
interactions from the class of diagrams shown in figure~\ref{fig-i1} for
two different wave functions: an adjoint field with a flavor wave-function
$\half\lambda_8$ involving only the first three flavors in the global
$SU(M)$; and a singlet field, with wave-function $\half I/\sqrt M$,
proportional to the identity in the $SU(M)$ flavor space (and thus of
course involving all the flavors). Both wave functions also contain the
usual $1/\sqrt N$ for color. Now in figure~\ref{fig-i1}, the color factors
are the same --- a factor of $N$ for the loop and three factor of $1/\sqrt
N$ from the wave-functions. Thus both interactions are suppressed by
$1/\sqrt N$ as expected. But the flavor factors are different. The adjoint
coupling is proportional to ${1\over8}\tr\lambda_8^3=\cO(1)$. But the
singlet interactions have an additional suppression of ${1\over8M^{3/2}}\tr
I=\cO(1/\sqrt M)$. This is a general feature of such states.

Thus in thinking about interactions, we can completely ignore singlets and
other states that have significant contributions from a number of flavors that
goes to infinity. This is a great convenience, because it means that we do not
have to worry about hidden flavor factors in wave-functions. They are only
there if the states are unimportant anyway. When we discuss the interactions
below, we will not often remind the reader of this --- we will simply assume
that we are dealing with states involving only a finite number of
flavors.\footnote{See, however, Appendix \ref{condensates}, for a discussion
of why internal loops cannot be ignored in the effective potential.} 
{\figsize\begin{figure}[htb]
$$\beginpicture
\setcoordinatesystem units <\tdim,\tdim>
\stpltsmbl
\barrow from 2 100 to -3 100
\barrow from -2 -100 to 3 -100
\barrow from 100 2 to 100 -3
\barrow from 20 -3 to 20 2
\circulararc 106.26 degrees from 60 80 center at 120 0
\setdashes
\circulararc 106.26 degrees from 60 -80 center at 0 0
\bsmbl
\setdots
\circulararc 253.74 degrees from 60 80 center at 0 0
\setshadesymbol <0pt,,0pt,> ({\tiny N} <0pt,0pt>)
\setshadegrid span <10\tdim>
\hshade -100 0 0 -90 -43.6 43.6 -80 -60.0 60.0 /
\hshade 80 -60.0 60.0 90 -43.6 43.6 100 0 0 /
\hshade -80 -60.0 60.0 -70 -71.4 48.6 -60 -80.0 40.0 -50 -86.6 33.4 -40 -91.7
28.3 -30 -95.4 24.6 -20 -98.0 22.0 -10 -99.5 20.5 0 -100. 20. 10 -99.5 20.5 20
-98.0 22.0 30 -95.4 24.6 40 -91.7 28.3 50 -86.6 33.4 60 -80.0 40.0 70 -71.4
48.6 80 -60.0 60.0 /\put {\Large$\bullet$} at -100 0
\setshadesymbol <0pt,,0pt,> ({\tiny M} <0pt,0pt>)
\hshade -80 60.0 60.0 -70 48.6 71.4 -60 40.0 80.0 -50 33.4 86.6 -40 28.3 91.7
-30 24.6 95.4 -20 22.0 98.0 -10 20.5 99.5 0 20. 100. 10 20.5 99.5 20 22.0 98.0
30 24.6 95.4 40 28.3 91.7 50 33.4 86.6 60 40.0 80.0 70 48.6 71.4 80 60.0 60.0
/
\put {$\clubsuit$} at 60 80
\put {$\clubsuit$} at 60 -80
\linethickness=0pt
\putrule from -150 0 to 150 0
\putrule from 0 -100 to 0 100
\endpicture$$
\caption{\figsize\sf\label{fig-i2} A class of diagrams contributing to
the interactions
of an $A\bar A$ adjoint mesons with a baryon-antibaryon pair in leading order
in large $N$.}\end{figure}}

Figure~\ref{fig-i1} is related to figure~\ref{fig-p2}. If we cut the diagrams
in figure~\ref{fig-p2} through the internal loop, the cut crosses two regions
that look like $A\bar A$ meson propagators. Thus the imaginary part will be
related to the partial width of the meson state into meson pairs. We see from
figure~\ref{fig-i1} that the individual couplings are small --- of order
$1/\sqrt N$. But because there are order $M$ states that can appear in the
final 2-meson state, the contribution to the width is order 1. This is what we
expect in a matrix model.

In a similar way, we can construct diagrams that contribute to the
interactions of baryons and mesons or multi-baryon states. For example, the
diagrams illustrated in figure~\ref{fig-i2} contribute to the interaction of
an $A\bar A$ adjoint mesons with a baryon-antibaryon pair in leading order
in large $N$. This is related to a cut through the internal loop in
figure~\ref{fig-p4}.

{\figsize\begin{figure}[htb]
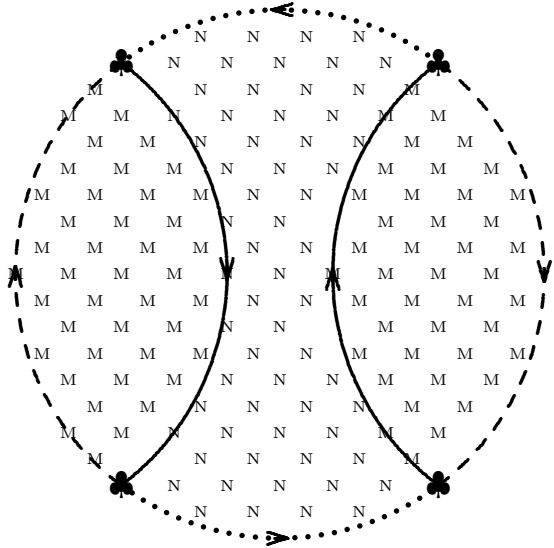

$$\beginpicture
\setcoordinatesystem units <\tdim,\tdim>
\stpltsmbl
\barrow from 2 100 to -3 100
\barrow from -2 -100 to 3 -100
\barrow from 100 2 to 100 -3
\barrow from -100 -2 to -100 3
\barrow from 20 -3 to 20 2
\barrow from -20 3 to -20 -2
\circulararc 106.26 degrees from 60 80 center at 120 0
\circulararc 106.26 degrees from -60 -80 center at -120 0
\setdashes
\circulararc 106.26 degrees from 60 -80 center at 0 0
\circulararc 106.26 degrees from -60 80 center at 0 0
\bsmbl
\setdots
\circulararc 72.74 degrees from 60 80 center at 0 0
\circulararc 72.74 degrees from -60 -80 center at 0 0
\setshadesymbol <0pt,,0pt,> ({\tiny N} <0pt,0pt>)
\setshadegrid span <10\tdim>
\hshade -100 0 0 -90 -43.6 43.6 -80 -60.0 60.0 /
\hshade 80 -60.0 60.0 90 -43.6 43.6 100 0 0 /
\hshade -80 -60.0 60.0 -70 -48.6 48.6 -60 -40.0 40.0 -50 -33.4 33.4 -40 -28.3
28.3 -30 -24.6 24.6 -20 -22.0 22.0 -10 -20.5 20.5 0 -20. 20. 10 -20.5 20.5 20
-22.0 22.0 30 -24.6 24.6 40 -28.3 28.3 50 -33.4 33.4 60 -40.0 40.0 70 -48.6
48.6 80 -60.0 60.0 /
\setshadesymbol <0pt,,0pt,> ({\tiny M} <0pt,0pt>)
\hshade -80 60.0 60.0 -70 48.6 71.4 -60 40.0 80.0 -50 33.4 86.6 -40 28.3 91.7
-30 24.6 95.4 -20 22.0 98.0 -10 20.5 99.5 0 20. 100. 10 20.5 99.5 20 22.0 98.0
30 24.6 95.4 40 28.3 91.7 50 33.4 86.6 60 40.0 80.0 70 48.6 71.4 80 60.0 60.0
/
\hshade -80 -60.0 -60.0 -70 -71.4 -48.6 -60 -80.0 -40.0 -50 -86.6 -33.4 -40
-91.7 -28.3 -30 -95.4 -24.6 -20 -98.0 -22.0 -10 -99.5 -20.5 0 -100. -20. 10
-99.5 -20.5 20 -98.0 -22.0 30 -95.4 -24.6 40 -91.7 -28.3 50 -86.6 -33.4 60
-80.0 -40.0 70 -71.4 -48.6 80 -60.0 -60.0 /
\put {$\clubsuit$} at 60 80
\put {$\clubsuit$} at 60 -80
\put {$\clubsuit$} at -60 -80
\put {$\clubsuit$} at -60 80
\linethickness=0pt
\putrule from -150 0 to 150 0
\putrule from 0 -100 to 0 100
\endpicture$$
\caption{\figsize\sf\label{fig-i3} A class of diagrams contributing to the
4-baryon interactions in leading order in large $N$.}
\end{figure}} 
In this model, we can show that the $N$ counting for the $ABC$ baryons is
actually the same as for the mesons. Each additional baryon in a graph
gives a factor of $1/\sqrt N$. To see this, note that because the baryon
propagator graphs like those in figure~\ref{fig-p5} have both an $N$ loop
and an $M$ loop, the wave-function renormalization factor for a baryon goes
like $1/\sqrt{N M}\sim 1/N$, smaller by $1/\sqrt N$ than for a
meson. However, this is compensated by the $B$ line that emerges from each
baryon vertex. This counts like a gluon line because it carries both and
$N$ and $M$ indices. Thus the extra $1/\sqrt N$ in the baryon wave function
acts like a gluon coupling that compensates the extra loop associated with
the $B$ line. Adjoint mesons and $ABC$ baryons both act like matrix fields
with couplings suppressed by $1/\sqrt N$.

{\figsize\begin{figure}[htb]
$$\beginpicture
\setcoordinatesystem units <\tdim,\tdim>
\stpltsmbl
\barrow from -2 100 to -3 100
\barrow from 2 -100 to 3 -100
\barrow from -2 -50 to -3 -50
\barrow from 2 50 to 3 50
\bsmbl
\setdots
\circulararc 360 degrees from 100 0 center at 0 0
\circulararc 360 degrees from 50 0 center at 0 0
\setshadesymbol <0pt,,,> ({\tiny N} <0pt,0pt>)
\setshadegrid span <10\tdim>
\vshade -100 0 0 -90 -43.6 43.6 -80 -60.0 60.0 -70 -71.4 71.4 -60 -80.0 80.0
-50 -86.6 86.6 /
\vshade 50 -86.6
86.6 60 -80.0 80.0 70 -71.4 71.4 80 -60.0 60.0 90 -43.6 43.6 100 0 0 /
\vshade -50 -86.6 0 -40 -91.7 -30.0 -30 -95.4 -40.0 -20 -98.0 -45.8 -10 -99.5
-49.0 0 -100. -50.0 10 -99.5 -49.0 20 -98.0 -45.8 30 -95.4 -40.0 40 -91.7
-30.0 50 -86.6 0 /
\vshade -50 0 86.6 -40 30.0 91.7 -30 40.0 95.4 -20 45.8 98.0 -10 49.0 99.5 0
50.0 100. 10 49.0 99.5 20 45.8 98.0 30 40.0 95.4 40 30.0 91.7 50 0 86.6 /
\put {\Large$\bullet$} at -70.7 -70.7
\put {\Large$\bullet$} at 70.7 -70.7
\put {\Large$\bullet$} at -70.7 70.7
\put {\Large$\bullet$} at 70.7 70.7
\linethickness=0pt
\putrule from -150 0 to 150 0
\putrule from 0 -100 to 0 100
\endpicture$$
\caption{\figsize\sf\label{fig-i4} A class of diagrams that
contribute to the interactions
of four $A\bar A$ adjoint mesons in leading order in large $N$. These graphs
contribute to 4-meson couplings of order $1/N$ (because $M/N^2=\cO(1/N)$).}
\end{figure}}
{\figsize\begin{figure}[htb]
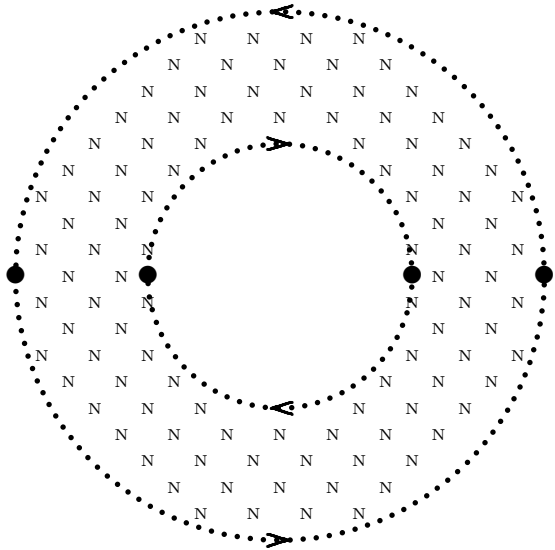

$$\beginpicture
\setcoordinatesystem units <\tdim,\tdim>
\stpltsmbl
\barrow from -2 100 to -3 100
\barrow from 2 -100 to 3 -100
\barrow from -2 -50 to -3 -50
\barrow from 2 50 to 3 50
\bsmbl
\setdots
\circulararc 360 degrees from 100 0 center at 0 0
\circulararc 360 degrees from 50 0 center at 0 0
\setshadesymbol <0pt,,,> ({\tiny N} <0pt,0pt>)
\setshadegrid span <10\tdim>
\vshade -100 0 0 -90 -43.6 43.6 -80 -60.0 60.0 -70 -71.4 71.4 -60 -80.0 80.0
-50 -86.6 86.6 /
\vshade 50 -86.6
86.6 60 -80.0 80.0 70 -71.4 71.4 80 -60.0 60.0 90 -43.6 43.6 100 0 0 /
\vshade -50 -86.6 0 -40 -91.7 -30.0 -30 -95.4 -40.0 -20 -98.0 -45.8 -10 -99.5
-49.0 0 -100. -50.0 10 -99.5 -49.0 20 -98.0 -45.8 30 -95.4 -40.0 40 -91.7
-30.0 50 -86.6 0 /
\vshade -50 0 86.6 -40 30.0 91.7 -30 40.0 95.4 -20 45.8 98.0 -10 49.0 99.5 0
50.0 100. 10 49.0 99.5 20 45.8 98.0 30 40.0 95.4 40 30.0 91.7 50 0 86.6 /
\put {\Large$\bullet$} at -100 0
\put {\Large$\bullet$} at -50 0
\put {\Large$\bullet$} at 100 0
\put {\Large$\bullet$} at 50 0
\linethickness=0pt
\putrule from -150 0 to 150 0
\putrule from 0 -100 to 0 100
\endpicture$$
\caption{\figsize\sf\label{fig-i5} A class of diagrams that do not
contribute to the interactions
of four $A\bar A$ adjoint mesons in leading order in large $N$. These graphs
contribute to 4-meson couplings of order $1/N^2$, because the internal loop
does not contribute a factor of $M$.}
\end{figure}}
Figure~\ref{fig-i3} illustrates graphs that contribute to 4-baryon
interactions. You
can check explicitly that the coupling is order $1/N$, in line with the
general argument in the previous paragraph.

In all these leading interaction diagrams, the flavor structure is
associated with a single flavor loop around the outside. Flavor loops
inside as in figure~\ref{fig-i4} are OK, but only because they involve a
sum over all flavors. If mesons or baryons couple to two separate loops, as
in figure~\ref{fig-i5}, the diagram is suppressed. Note that this implies
that there is no coupling between $A\bar A$ and $C\bar C$ adjoint mesons in
leading order in $N$ except in diagrams involving the baryons on the
external line, even though both kinds of meson have leading interactions
with the $ABC$ baryons.

Before we go on to more complicated and interesting models, it is worth
noting that the large $N$ limit that we have discussed in this and the
previous section is completely consistent with our assumption that there is
no phase transition in this model as a function of the ratio of gauge
couplings. The fermion bound states that exist in large $N$ are precisely
those required to be massless to saturate the anomaly condition. And the
quantum numbers of the massless fermion states match precisely with what we
find in the Higgs picture that emerges if one coupling is much larger than
the other.\footnote{This will be elaborated in the next section.} The
$A\bar A$ and $C\bar C$ mesons that exist in large $N$ are expected to play
no role in global symmetry breaking and there is no reason to expect any of
them to be light.

\section{$k\neq0$\label{knotzero}}

Let us now consider the field theory associated with the moose shown in
figure~\ref{fig-2} in the limit $N,M\rightarrow\infty$ with $M/N$ and $k$
fixed. Define the dimensionless variable
\begin{equation}
\ximn\equiv\Lambda_M/\Lambda_N
\label{ximn}
\end{equation}
In this theory, we know that there is a phase transition as a function
of $\ximn$ because the theory has different symmetry and
particle content in the two limits 
$\ximn\gg1$
and
$\ximn\ll1$. To see this explicitly, let us label the fields and
symmetries as shown in figure~\ref{fig-k1}
{\figsize\begin{figure}[htb]
$$\beginpicture
\setcoordinatesystem units <\tdim,\tdim>
\stpltsmbl
\circulararc 360 degrees from 215 0 center at 200 0
\circulararc 360 degrees from 115 0 center at 100 0
\put {$M$+$k$~~} at 0 0
\put {$M$} at 200 0
\put {$N$} at 100 0
\put {$k$} at 100 100
\put {$N$} at 300 0
\tarrow from 100 15 to 100 50
\plot 100 50 100 85 /
\put {$A$} at 50 -15 
\put {$B$} at 150 -15 
\put {$C$} at 250 -15 
\put {$D$} at 85 50 
\tarrow from 15 0 to 50 0
\plot 50 0 85 0 /
\tarrow from 115 0 to 150 0
\plot 150 0 185 0 /
\tarrow from 215 0 to 250 0
\plot 250 0 285 0 /
\setdashes
\put {$SU(M$+$k)$} at 0 -50
\tarrow from 0 -40 to 0 -10  
\put {$SU(k)$} [r] at 55 100
\tarrow from 60 100 to 90 100
\put {$SU(M)_g$} at 200 -50
\tarrow from 200 -40 to 200 -20  
\put {$SU(N)$} at 300 -50
\tarrow from 300 -40 to 300 -10  
\put {$SU(N)_g$} at 100 -50
\tarrow from 100 -40 to 100 -20  
\endpicture$$
\caption{\figsize\label{fig-k1}} \sf The Moose shown in
figure~\protect\ref{fig-2} with the fermion fields labeled.
\end{figure}}

For $\ximn\gg1$, the $SU(N)_g$ gauge coupling is small at the $SU(M)$
confinement scale, $\Lambda_M$. Thus we can analyze the symmetry breaking
produced by the $SU(M)_g$ gauge theory ignoring the effect of the $SU(N)$
interactions.  We therefore expect the $B$ and $C$ fermions to condense,
breaking the $SU(N)_g\times SU(N)$ symmetry down to a diagonal
$SU(N)$~\cite{vafawitten}. In the process, the Goldstone bosons associated
with the symmetry breaking are eaten by the Higgs mechanism, leaving the
$SU(N)_g$ gauge bosons, $G_N$, massive, but light because the gauge
coupling is small. This process leaves the global $SU(M$+$k)\times SU(k)$
symmetry unbroken, and leaves the $A$ and $D$ fields in the theory in the
infrared as free massless fermions.\footnote{That is, they interact only
through nonrenormalizable interactions.}

Summarizing, 
\begin{equation}
\fbox{\begin{tabular}{c}
for $\ximn\gg1$ \\
{\bf the unbroken global symmetry} in the theory is\\
$SU(M$+$k)\times SU(k)\times SU(N)$\\
{\bf the light particles} in the theory are:\\
massless LH fermions $A$ transforming like $(M$+$k,1,\bar N)$;\\
massless LH fermions $D$ transforming like $(1,\bar k,N)$;\\
light gauge bosons $G_N$ transforming like $(1,1,N^2$$-$$1)$.\\
\end{tabular}}
\label{ximngg}
\end{equation}

For $\ximn\ll1$, the $SU(M)_g$ gauge coupling is small at the $SU(N)$
confinement scale, $\Lambda_N$. Then we can analyze the symmetry breaking
produced by the $SU(N)_g$ gauge theory ignoring the effect of the $SU(M)$
interactions.  We therefore expect the $A$ fermions to condense with the
$B$ and $D$ fermions.  If we could completely ignore the $SU(M)_g$ gauge
couplings, the $B$ and $D$ fermions would combine into $M$+$k$ fermions
with a global $SU(M$+$k)$ symmetry. In this limit, we would have an
$SU(M$+$k)\times SU(M$+$k)$ symmetry spontaneously broken down to
$SU(M$+$k)$, producing an $SU(M$+$k)$ adjoint of Goldstone bosons. However,
the $SU(M)_g$ gauge interactions break this down to $SU(M)\times SU(k)$ and
eat an $SU(M)$ adjoint of Goldstone bosons, leaving the $SU(M)_g$ gauge
bosons, $G_M$, massive, but light because the gauge coupling is small. The
$SU(N)$ symmetry is unbroken, and the $C$ fermions are still massless,
transforming like $(M,1,\bar N)$ under the $SU(M)\times SU(k)\times SU(N)$.
The uneaten Goldstone bosons transform like
$(1,k^2$$-1$$,1)\oplus(1,1,1)\oplus(\bar M,k,1)\oplus(M,\bar k,1)$
corresponding to the coset space
\begin{equation}
{
SU(M\mbox{+}k)\times SU(k)
\over
SU(M)\times SU(k)
}
\label{gbs}
\end{equation}

Summarizing,
\begin{equation}
\fbox{\begin{tabular}{c}
for $\ximn\ll1$ \\
{\bf the unbroken global symmetry} in the theory is\\
$SU(M)\times SU(k)\times SU(N)$\\
{\bf the light particles} in the theory are:\\
massless LH fermions $C$ transforming like $(M,1,\bar N)$;\\
light gauge bosons $G_M$ transforming like $(M^2$$-$$1,1,1)$;\\
Goldstone bosons in the coset space\\
$\displaystyle{
SU(M\mbox{+}k)\times SU(k)
\over
SU(M)\times SU(k)
}$
\end{tabular}}
\label{ximnll}
\end{equation}

Evidently, (\ref{ximngg}) and (\ref{ximnll}) are different, so there must a
phase transition between these two regions. The question that we want to
address is --- What kind is it? We will show that the transition is exactly
what we would expect from a second order transition caused by the
expectation value of an $A$-$D$ bound state meson. This is a nontrivial
consistency check because the $A$-$D$ mesons are among the (few) states
that we expect in the large $N$ limit.

To see that this works in detail, let us look in detail at the region
$\ximn\approx1$, but assuming that the global chiral symmetries are
unbroken --- that is that we are in a phase continuously connected to
(\ref{ximngg}) for $\ximn\gg1$.

We expect fermion bound states in two channels: $A$-$B$-$C$ bound states
which transform as $(M$+$k,1,\bar N)$, and a $\bar C$-$\bar B$-$D$ bound
states which transform as $(1,\bar k,N)$.  The anomaly conditions are
saturated simply if we assume that the lightest state in each channel is a
massless LH fermion, so that is what we will assume. {\bf That it is
possible to saturate the anomaly conditions with the states we expect in
large $N$ is a nontrivial check on the assumption that we are in a phase
continuously connected to (\ref{ximngg}).} Note further that the anomaly
conditions determine uniquely whether the massless states are left handed
or right handed. We can usefully write the interpolating fields for these
massless states as matrix fields,
\begin{equation}
\begin{tabular}{c}
$\psi_{ABC}$ is an $M$+$k\times N$ matrix;\\
$\psi_{\bar C\bar BD}$ is an $N\times k$ matrix.
\end{tabular}
\label{fmatrixfields}
\end{equation}
The $A$-$D$ bound state meson can also be described by a matrix field,
\begin{equation}
\mbox{$\phi_{AD}$ is an $M$+$k\times k$ matrix.}
\label{smatrixfield}
\end{equation} 
In terms of these matrix fields, we can write the leading Yukawa coupling
between these fermions as
\begin{equation}
\tr\,\left(\psi_{\bar C\bar BD}\,\phi_{AD}^\dagger\, \psi_{ABC}\right)
\label{yukawa}
\end{equation}

{\bf Note that one reason we have singled out the $A$-$D$ bound state is
that it is the only scalar that has a renormalizable Yukawa coupling in
leading order in large $N$.} In the $k=0$ theory, where the $D$ field does
not exist, there are no renormalizable Yukawa couplings at all. This is
because the massless fermions in this model are all left-chiral and carry a
conserved $U(1)$ quantum number.  The $U(1)$ symmetry means that the
massless fermions $\psi_L$ must appear in the combination $\bar\psi_L
\psi_L$, which cannot be used to construct a Lorentz invariant dimension
four Yukawa coupling. Of course, the other reason that we have singled out
this field is that it is implicated in the phase transition in large $N$,
as we will now show more explicitly.

The phase for $\ximn\ll1$ corresponds to a large negative mass squared for
the $A$-$D$ mesons. The mesons transform like $(M$+$k,\bar k,1)$ under the
global symmetries, described by an $M$+$k\times k$ matrix.  If $\phi_{AD}$
develops an expectation value of the form
\begin{equation}
\langle\phi_{AD}\rangle\propto\pmatrix{
I\cr
0\cr}\label{advev}
\end{equation}
where $I$ is the $k\times k$ identity matrix, it breaks the symmetry down
to $SU(M)\times SU(k)\times SU(N)$ producing the Goldstone bosons of
(\ref{gbs}). The non-Goldstone components of $\phi_{AD}$ get mass due to
leading self-couplings of $\phi_{AD}$, but they are light if the
expectation value is small. The fermion field $\psi_{ABC}$ breaks up into
an $(M,1,\bar N)\oplus(1,k,\bar N)$. At the same time, the Yukawa
couplings, (\ref{yukawa}), gives a small mass to the fermions that
transform nontrivially under $SU(k)$, leaving only the $(M,1,\bar N)$ of
massless fermions.  If the transition is second order, the complete
$\phi_{AD}$ multiplet that exists as $\ximn$ approaches 1 from above gets
lighter as the transition is approached, and the sign of mass squared
changes at the transition, leading to the expectation value (\ref{advev}).

Summarizing,
\begin{equation}
\fbox{\begin{tabular}{c}
for $\ximn$ not much greater than 1\\
{\bf the unbroken global symmetry} in the theory is\\
$SU(M$+$k)\times SU(k)\times SU(N)$\\
{\bf the light particles} in the theory are:\\
massless LH fermions $\psi_{ABC}$ transforming like $(M$+$k,1,\bar N)$;\\
massless LH fermions $\psi_{\bar C\bar BD}$ transforming like $(1,\bar
k,N)$;\\
light scalar bosons $\phi_{AD}$ transforming like $(M$+$k,\bar k,1)$.\\
\end{tabular}}
\label{ximnone}
\end{equation}

Let us now think more carefully about whether and how the putative second
order phase transition for $\ximn\approx1$ in (\ref{ximnone}) matches
smoothly onto the our reliable calculations of the particle content in the
extreme regions, (\ref{ximngg}) and (\ref{ximnll}).  As $\ximn$ falls from
a large value, the unbroken global symmetries and the massless fermions
match trivially without any physical change. There is a change in our {\bf
interpretation} of the nature of the global symmetries and the fermions,
but this is in the eye of the beholder. The $G_N$ gauge bosons in
(\ref{ximngg}) get heavier and heavier and eventually disappear into the
hadronic mess at the confinement scale. They have the quantum numbers of an
$SU(N)$ $C\bar C$ bound state, and look more and more like a strongly
interacting bound state as $\ximn$ decreases.  The $\phi_{AD}$ is more
interesting. We can imagine two ways in which this state might arise as
$\ximn$ falls. It might emerge from the hadronic mess and get lighter with
decreasing $\ximn$. Or it might be plucked from the $A$-$D$ fermion-fermion
continuum when some interaction in the effective theory becomes
sufficiently strong. In this instance, the former picture is not
supportable for large $\ximn$. The hadronic states that are strongly
interacting for large $\ximn$ do not carry any $SU(M$+$k)\times SU(k)$
quantum numbers, and thus the $\phi_{AD}$ cannot be found there for large
$\ximn$. Either the $\phi_{AD}$ state only develops as $\ximn$ approaches 1,
where the strong interactions involve all the fermions, or strong
interactions must develop at intermediate $\ximn$ to bind the meson state
at some value of $\ximn$. There is an obvious candidate for these strong
interactions. The exchange of the massive $G_N$ gauge bosons produce an
attractive force in the $A$-$D$ channel. This is the original inspiration
for the physics of
topcolor~\cite{topcolor}. At low energies, $G_N$ exchange
looks like the 4-fermion
interaction of the NJL model~\cite{njl} but it is cut off at the mass of
$G_N$, which for large $\ximn$ is much smaller than the cut-off in the
effective theory, which is of order $\Lambda_M$. In this situation, the
attractive force is weak, and the $\phi_{AD}$ is not bound.  The second
order picture of the phase transition makes sense if when $G_N$ gets
sufficiently heavy, the interaction becomes strong enough to bind the
meson. This is plausible, and provides an interesting commentary on the NJL
picture. Of course, it is not obvious that these two pictures of the
$\phi_{AD}$, emerging from the hadronic mess and being bound from the
continuum, are fundamentally different. If the binding does not take place
until $\ximn\approx1$, these two may simply be complementary descriptions
of the same physics~\cite{sekhar}. It may be important for topcolor model
building to understand these issues better.

Below the transition, as $\ximn$ gets smaller and smaller, again the
massless fermions match trivially. The massive fermions and scalars get
heavier and heaver and approach and merge into the hadronic mess (now the
quantum numbers are right to make this sensible). The $G_M$ gauge boson,
with the quantum numbers of an $A\bar A$ bound state, emerges from the
hadronic mess and gets lighter and lighter.

One more thing is required to show the consistency of the second order
phase transition in the large $N$ limit. We must show that the symmetry
breaks in the right way when the sign of the $\phi_{AD}$ mass squared term
changes.

It follows from our assumption about the $k=0$ theory that the Swiss cheese
slices in the leading order graphs are flavor blind. As we will show in
more detail in appendix~\ref{condensates}, we can't prove this. We can't
prove that condensates involving an infinite number of quarks don't affect
the Swiss cheese slices. But we have assumed that it is true. It is part of
our assumption about the $k=0$ theory. We know that no such condensates
appear in the extreme limits of the theory (from the Vafa-Witten
argument~\cite{vafawitten}).\footnote{We know from large $N$ arguments that
condensates involving only a small number of flavors do not appear, but the
large $N$ arguments show that they don't matter anyway.} And our smoothness
assumption thus implies that do not appear for intermediate $\ximn$

This plus standard large $N$ arguments shows that non-zero $k$ doesn't
change the Swiss cheese slices in leading order in $N$.  This is probably
obvious. Internal loops are suppressed by powers of $N$. They have leading
effects only if these are compensated by flavor sums. But $k$ is finite, so
the contributions of the extra quarks to internal flavor loops is
nonleading in $N$.

Because the Swiss cheese slices have no flavor structure, the flavor
structure of the leading particle interactions is associated with the
flavor loop around the outside.  Thus if we write an effective field theory
in the neighborhood of the phase transition, the effective potential has
the Coleman-Witten form in leading order in $N$~\cite{cwit}, in which the
only leading terms involve singles traces of products of the matrix fields,
$\phi_{AD}$, $\psi_{ABC}$ and $\psi_{\bar C\bar BD}$.  The parameters in
the effective field theory will depend on $\ximn$. We have seen that this
is consistent with a second order phase transition. All that is required
for a second order phase transition is that the mass parameter go through
zero for some value of $\ximn$.

Generically the second order phase transition will preserve the $SU(k)$
symmetry because of the Coleman-Witten form. In leading order, the
potential in the effective low energy theory is
\begin{equation}
{\lambda\over4}\,\tr\left(\phi_{AD}^\dagger\,
\phi_{AD}\,\phi_{AD}^\dagger\,\phi_{AD}\right)
+{m^2\over2}\,\tr\left(\phi_{AD}^\dagger\,\phi_{AD}\right)
\label{potential}
\end{equation}
This can be written as
\begin{equation}
{\lambda\over4}\,\tr\left[\left(\phi_{AD}^\dagger\, \phi_{AD}+\mu^2\,I
\right)^2\right]+\mbox{constant}
\label{potential2}
\end{equation}
where
\begin{equation}
\mu^2=m^2/\lambda
\label{potential3}
\end{equation}
and $I$ is the $k\times k$ identity. When $m^2$ is negative, this is
obviously
minimized for
\begin{equation}
\phi_{AD}^\dagger\, \phi_{AD}=-\mu^2\,I
\label{potential4}
\end{equation}
which implies that vacuum value can be rotated into the form (\ref{advev}).
This is necessary for the consistency of the whole scheme, and again it
comes out automatically.

Thus there is a plausible picture of a completely smooth evolution of the
theory from one extreme to the other as a function of $\ximn$. The phase
transition can be second order.  We regard the arguments in this section as
strong evidence that the phase transition is second order for finite $k$ in
the large $N$ limit.  We must admit that had hoped to find even stronger
results. The fact that the Swiss cheese slices that are the background in
which all the fermions propagate are independent of $k$ is very
suggestive. It certainly means that the there is no evidence of a phase
transition in the vacuum energy in leading order. However, we cannot rule
out the possibility that a first order transition takes place that only
involves an infinitesimal fraction of the fermions and does not have a
leading effect on the vacuum energy.

\section{$k,\ell\neq0$\label{klnotzero}}

Let us now perform a similar analysis of the theory shown in
figure~\ref{fig-3}. Again, we begin by labeling the fields and symmetries
as shown in figure~\ref{fig-kl1}
{\figsize\begin{figure}[htb]
$$\beginpicture
\setcoordinatesystem units <\tdim,\tdim>
\stpltsmbl
\circulararc 360 degrees from 215 0 center at 200 0
\circulararc 360 degrees from 115 0 center at 100 0
\put {$M$+$k$~~} at 0 0
\put {$M$} at 200 0
\put {$N$} at 100 0
\put {$k$} at 100 100
\put {~$N$+$\ell$} at 300 0
\put {$\ell$} at 200 100
\tarrow from 200 85 to 200 50
\plot 200 50 200 15 / 
\tarrow from 100 15 to 100 50
\plot 100 50 100 85 /
\put {$A$} at 50 -15 
\put {$B$} at 150 -15 
\put {$C$} at 250 -15 
\put {$D$} at 85 50 
\put {$E$} at 215 50
\tarrow from 15 0 to 50 0
\plot 50 0 85 0 /
\tarrow from 115 0 to 150 0
\plot 150 0 185 0 /
\tarrow from 215 0 to 250 0
\plot 250 0 285 0 /
\setdashes
\put {$SU(M$+$k)$} at 0 -50
\tarrow from 0 -40 to 0 -10  
\put {$SU(k)$} [r] at 55 100
\tarrow from 60 100 to 90 100  
\put {$SU(\ell)$} [l] at 245 100
\tarrow from 240 100 to 210 100  
\put {$SU(M)_g$} at 200 -50
\tarrow from 200 -40 to 200 -20  
\put {$SU(N$+$\ell)$} at 300 -50
\tarrow from 300 -40 to 300 -10  
\put {$SU(N)_g$} at 100 -50
\tarrow from 100 -40 to 100 -20  
\endpicture$$
\caption{\figsize\label{fig-kl1}} \sf The Moose shown in
figure~\protect\ref{fig-3} with the fermion fields labeled.
\end{figure}}

The arguments for the two limiting cases, $\ximn\gg1$ and $\ximn\ll1$ should
by now be familiar, so we will simply summarize the results. 
\begin{equation}
\fbox{\begin{tabular}{c}
for $\ximn\gg1$ \\
{\bf the unbroken global symmetry} in the theory is\\
$SU(M$+$k)
\times SU(k)
\times SU(\ell)
\times SU(N)$\\
{\bf the light particles} in the theory are:\\
massless LH fermions $A$ transforming like $(M$+$k,1,1,\bar N)$;\\
massless LH fermions $D$ transforming like $(1,\bar k,1,N)$;\\
light gauge bosons $G_N$ transforming like $(1,1,1,N^2$$-$$1)$.\\
Goldstone bosons in the coset space\\
$\displaystyle{
SU(N\mbox{+}\ell)\times SU(\ell)
\over
SU(N)\times SU(\ell)
}$
\end{tabular}}
\label{ximngg2}
\end{equation}

\begin{equation}
\fbox{\begin{tabular}{c}
for $\ximn\ll1$ \\
{\bf the unbroken global symmetry} in the theory is\\
$SU(M)
\times SU(k)
\times SU(\ell)
\times SU(N$+$\ell)$\\
{\bf the light particles} in the theory are:\\
massless LH fermions $C$ transforming like $(M,1,1,\bar{N\mbox{+}\ell})$;\\
massless LH fermions $E$ transforming like $(\bar M,1,\ell,1)$;\\
light gauge bosons $G_M$ transforming like $(M^2$$-$$1,1,1,1)$;\\
Goldstone bosons in the coset space\\
$\displaystyle{
SU(M\mbox{+}k)\times SU(k)
\over
SU(M)\times SU(k)
}$
\end{tabular}}
\label{ximnll2}
\end{equation}

The arguments of the previous section suggest that we should expect the
intermediate region, $\ximn\approx1$, to be continuously connected to each
of the extreme regions. The argument is very similar to what we have
already seen, but considerably more involved. We consider the light bound
states in the confining picture, assuming that the global symmetries are
unbroken, or perhaps broken only the vacuum values of composite fields.
For $\ximn\approx1$, we expect massless LH fermion bound states in each of the
four channels that are allowed in large $N$:
\begin{itemize}
\item
an $A$-$B$-$C$ bound state which transform as $(M$+$k,1,1,\bar
{N\mbox{+}\ell})$,
\item
a $\bar C$-$\bar B$-$D$ bound state which transform as $(1,\bar
k,1,N\mbox{+}\ell)$, 
\item
an $E$-$\bar B$-$\bar A$ bound state which transforms
as
$(\bar{M\mbox{+}k},1,\ell,1)$, 
\item
an $\bar D$-$B$-$\bar E$ bound state which
transforms as $(1,k,\bar \ell,1)$.
\end{itemize}
As before, this is the unique choice that saturates the
anomaly conditions.
Again, we can
usefully write the interpolating fields for these massless states as matrix
fields,
\begin{equation}
\begin{tabular}{c}
$\psi_{ABC}$ is an $M$+$k\times N$+$\ell$ matrix;\\
$\psi_{\bar C\bar BD}$ is an $N$+$\ell\times k$ matrix;\\
$\psi_{E\bar B\bar A}$ is an $\ell\times M$+$k$ matrix;\\
$\psi_{\bar DB\bar E}$ is an $k\times\ell$ matrix.
\end{tabular}
\label{fmatrixfields2}
\end{equation}

We also expect light bound state mesons,
$A$-$D$ and $E$-$C$. These
can also be described by matrix fields,
\begin{equation}
\begin{tabular}{c}
$\phi_{AD}$ is an $M$+$k\times k$ matrix;\\
$\phi_{EC}$ is an $\ell\times N$+$\ell$ matrix.
\end{tabular}
\label{smatrixfield2}
\end{equation} 
In terms of these matrix fields, we can write the leading Yukawa coupling
between these fermions as
\begin{equation}
\begin{array}{c}
\tr\,\left(\psi_{\bar C\bar BD}\,\phi_{AD}^\dagger\, \psi_{ABC}\right)\\
\tr\,\left(\psi_{E\bar B\bar A}\,\phi_{AD}\, \psi_{\bar DB\bar E}\right)\\
\tr\,\left(\psi_{ABC}\,\phi_{EC}^\dagger\, \psi_{E\bar B\bar A}\right)\\
\tr\,\left(\psi_{\bar DB\bar E}\,\phi_{EC}\, \psi_{\bar C\bar BD}\right)
\end{array}
\label{yukawa2}
\end{equation}

Summarizing,
\begin{equation}
\fbox{\begin{tabular}{c}
for $\ximn\approx 1$\\
{\bf the unbroken global symmetry} in the theory is\\
$SU(M$+$k)
\times SU(k)
\times SU(\ell)
\times SU(N$+$\ell)$\\
{\bf the light particles} in the theory are:\\
massless LH fermions $\psi_{ABC}$ transforming like $(M$+$k,1,1,\bar
{N\mbox{+}\ell})$;\\
massless LH fermions $\psi_{\bar C\bar BD}$ transforming like $(1,\bar
k,1,N$+$\ell)$;\\
massless LH fermions $\psi_{E\bar B\bar A}$ transforming like
$(\bar{M\mbox{+}k},1,\ell,1)$;\\
massless LH fermions $\psi_{\bar DB\bar E}$ transforming like $(1,
k,\bar\ell,1)$.\\
light scalar bosons $\phi_{AD}$ transforming like $(M$+$k,\bar k,1)$.\\
light scalar bosons $\phi_{EC}$ transforming like
$(1,1,\ell,\bar{N\mbox{+}\ell})$.
\end{tabular}}
\label{ximnone2}
\end{equation}

Now let us see how this goes smoothly into the two extreme limits. The
situation is fairly symmetrical, so we will just discuss $\ximn\gg1$. This
corresponds to positive mass squared for $\phi_{AD}$ and a negative mass
squared for $\phi_{EC}$, which develops a vacuum value of the form
\begin{equation}
\langle\phi_{EC}\rangle\propto\pmatrix{
I&0\cr
}
\label{advac2}
\end{equation}
where $I$ is the $\ell\times\ell$ identity matrix. This breaks the symmetry
down to $SU(M$+$k)
\times SU(k)
\times SU(\ell)
\times SU(N)$, and produces the appropriate Goldstone bosons. Under this
symmetry, the fermion fields break up as follows:
\begin{equation}
\begin{array}{c}
\psi_{ABC} 
: (M\mbox{+}k,1,1,\bar{N\mbox{+}\ell})
\rightarrow (M\mbox{+}k,1,1,\bar{N})\oplus
(M\mbox{+}k,1,\bar{\ell},1);\\
\psi_{\bar C\bar BD} 
: (1,\bar k,1,N\mbox{+}\ell)
\rightarrow (1,\bar k,1,N)\oplus
(1,\bar k,\ell,1);\\
\psi_{E\bar B\bar A} 
: (\bar{M\mbox{+}k},1,\ell,1)
\rightarrow (\bar{M\mbox{+}k},1,\ell,1);\\
\psi_{\bar DB\bar E} 
: (1,k,\bar\ell,1)
\rightarrow (1,k,\bar\ell,1)
\end{array}
\label{breakup}
\end{equation}
All of the fields that transform nontrivially under the $SU(\ell)$
get mass due to the Yukawa couplings, (\ref{yukawa2}). The massless fermions
are just what we expect from (\ref{ximngg2}).

Thus again, this is consistent with a second order phase transition --- or
rather, with {\bf two} second order phase transitions. There is no reason
for the mass squares of $\phi_{AD}$ and $\phi_{EC}$ to go to zero at the
same value of $\ximn$. We do expect that for $\ximn\approx1$, the theory
can be described by an effective field theory with both scalars
present. But our methods do not tell us which phase transition occurs
first. There may be a region near $\ximn=1$ in which both mass squares are
positive and the full symmetry is unbroken. Or there may be a region in
which both mass squares are negative, and the symmetry is broken down to
$SU(M)\times SU(k) \times SU(\ell)\times SU(N)$. We cannot say with the
tools we have used here.  In either case, however, theories with this
structure open up new possibilities for topcolor model building that are
only present with dynamical topcolor breaking.

\section{Conclusions}

We have analyzed a family of chiral gauge theories in the large-$N$ limit.
Our aim has been to develop new tools that may be useful in constructing
models with dynamically broken topcolor.  

The interactions in these models are quite simple in the large-$N$ limit:
couplings of mesons and baryons are given by planar diagrams with all
interpolating fields on the outermost quark loop.  Since the number of
flavors is large, the diagrams have a Swiss cheese-like structure, with
arbitrary numbers of internal quark loops forming holes in the diagram.
Couplings of baryons and mesons are suppressed by the usual factors of
$1/\sqrt{N}$.  The analysis is simplified by the fact that, contrary to
what one might expect, states involving an infinite number of flavors
decouple in the large $N$ limit.

The phase structure of these models has proven more difficult
to analyze.  We
have argued that models with $k,\ell\neq 0$ exhibit a simple pattern of
chiral symmetry breaking.  With the assumption that the $k=\ell=0$ model
does not break its global $SU(M)\times SU(N)$ symmetry, we have shown that
the model with $k\neq 0$ has two phases, one with $SU(M+k)\times SU(k)
\times SU(N)$ symmetry, and another where condensates break the symmetry to
$SU(M)\times SU(k) \times SU(N)$.  
We used a generalization of the Coleman-Witten
argument to show that the $SU(k)$ symmetry does not break.
Similar considerations apply to the more complicated case with $k,\ell \neq
0$, where we expect two phase transitions.

For these models to be useful for electroweak symmetry breaking, it is
crucial that their phase transitions are second order, or at least weakly
first order.  We have been unable to conclusively prove that this is the
case, but the hypothesis that the phase transitions are second order is
consistent with all the available evidence.  The effective theories contain
obvious candidates for symmetry breaking composite Higgs fields, and simple
assumptions about the low energy Higgs potential imply that the phase
transitions are indeed second order.  Hence it seems likely that these
models are relevant to topcolor models of
dynamical electroweak symmetry breaking.

Although we have been able to analyze many aspects of these models, our
work leaves a number of interesting questions unanswered.  In particular,
questions about the phase structure of these theories are both interesting
and very difficult.  It would be very exciting if more definitive
conclusions could be drawn. However, it seems reasonable to use models
of this kind for topcolor model building. One important feature of these
models (already noted in reference \cite{seesaw2}) is that the fields that
break the topcolor gauge symmetry also contribute to anomaly matching. The
richer structure of the models with $k$ and $\ell$ both non-zero, however,
are virgin territory, yet to be explored by model builders.

\newpage
\appendix
\section{Confusing condensates \label{condensates}}

There is something that we found confusing about condensates in the large
$N$ limit.  Our confusion results from the following
apparent paradox: The diagrams
that dictate the leading interactions in the model are not the same as those
that dictate the vacuum structure of the model.  As discussed in the text,
scattering of mesons and baryons in the model is given in leading order by
Swiss cheese slices where the external fields couple only to the outermost
quark loop.  Diagrams where the external fields couple to quark lines
enclosing a hole in the Swiss cheese are suppressed by the usual $1/N$
factor, and this $1/N$ factor is not compensated by a sum over flavors.  On
the other hand, the vacuum energy is determined by the sum over Swiss
cheese slices with no external lines.  If we consider a diagram with
insertions of some order parameter (such as, for instance, a quark
bilinear), then we must perform a trace on the flavor indices to get the
contribution to the vacuum energy.  The trace on flavor indices means that
the order parameters can be inserted on any quark line in the Swiss cheese.
There is no $1/N$ suppression of those diagrams where the order parameter
lies on a hole in the Swiss cheese, since the $1/N$ is compensated by the
sum over flavors.  So for the diagrams that determine the leading
interactions, the flavor structure resides only on the outermost quark
loop, while for the vacuum energy, the flavor structure can lie on any
quark loop.

To add to the confusion, we have argued that states involving an infinite
number of flavors decouple in the large $N$ limit.  However, when we
consider condensates, it is possible that the vacuum values of such fields
become larger with $N$ and so do not decouple from the vacuum structure of
the theory.  It is possible that if an instability develops in such a
direction, it is not stabilized by the leading order interactions, and
instead is driven to a large value, requiring us to go beyond leading order
in $N$ (or more precisely, to include the flavor structure of internal
loops in leading order) to determine the vacuum structure.

The presence of the subleading contributions from internal quark loops in
the vacuum energy complicates the analysis of the symmetry breaking.  If it
were not for these subleading contributions, the Coleman-Witten arguments
could be used to show that the $SU(N)\times SU(M)$ global symmetry does not
break.  However, in our case, we can construct simple examples that
preserve the flavor symmetry, and other examples which do not.  The vacuum
energy has the form
\begin{equation}
V(\Phi) \sim \sum \frac{\prod_{i=1}^{\ell} {\rm Tr}\,
\Phi^{n_i}}{N^{\ell-1}},
\end{equation}
where $\Phi$ is an order parameter with the flavor quantum numbers of
$A\bar{A}$ or $C\bar{C}$.  Each trace comes from a separate quark loop, and
the powers of $1/N$ come from the usual suppression of internal quark
loops.  Points of unbroken symmetry have $\Phi = \beta I$ for some $\beta$,
possibly zero.  Suppose the function $V(\beta I)$ has a minimum for some
value $\beta = \beta_*$.  Then the point $\Phi = \beta_* I$ is a stationary
point of the potential.  If we could show that this point is a minimum, then
we would have established that there is at least a local minimum with
unbroken symmetry.  However, there are simple (if highly non-generic)
examples where the stationary point is a saddle point rather than a
minimum.  For instance, take the case $N=M$, and consider the potential
\begin{equation}
V = -\frac{5}{8} m^2 {\rm Tr}\, \Phi^2 + \frac{1}{2 N} m^2 ({\rm Tr}\, \Phi)^2
+\frac{1}{16} {\rm Tr}\, \Phi^4.
\end{equation}
This has a single stationary point at $\beta_* = m$.  But by considering small
variations in the diagonal elements of $\Phi$ in the neighborhood of this
point, it can be shown that this is a saddle point, even as $N\rightarrow
\infty$.

We feel it is unlikely that the $SU(N)\times SU(M)$ flavor symmetry is
broken by any condensates.  We suspect that the only symmetry-breaking
condensates are those that occur in the models with finite $k$ and $l$, and
that the low energy theory always preserves (at least) the $SU(N)\times
SU(M)$ flavor symmetry.  However we have been unable to prove this in
general, and the considerations of this appendix seem to indicate that
large $N$ and anomaly matching are insufficient to establish the pattern of
symmetry breaking.

\end{document}